# Performance Evaluation of IPTV over WiMAX Networks Under Different Terrain Environments

Jamil M. Hamodi, Ravindra C. Thool
Information Technology Department,
Shri Guru Gobind Singhji Institute of Engineering and Technology (SGGS)
Swami Ramanand Teerth Marathwada University (SRTM)
Nanded, INDIA

**Abstract:-** Deployment Video on Demand (VoD) over the next generation (WiMAX) has become one of the intense interest subjects in the research these days, and is expected to be the main revenue generators in the near future. In this paper, the performance of Quilty of Service of video streaming (IPTV) over fixed mobile WiMax network is investigated under different terrain environments, namely Free Space, Outdoor to Indoor and Pedestrian. OPNET is used to investigate the performance of VoD over WiMAX. Our findings analyzing different network statistics such as packet lost, path loss, delay, network throughput.

## I. INTRODUCTION

Worldwide Interoperability for Microwave Access (WiMAX) technology is the only wireless system capable of offering high QoS at high data rates for IP networks. The high data rate and Quality of Service (QoS) assurance provided by this standard has made it commercially viable to support multimedia applications such as video telephony, video gaming, and mobile TV broadcasting. System architecture to support high definition video broadcasting (like H.264/AVC and SVC) that provides a mobility of 30 kmph in an urban and sub-urban environment has been developed. Some aspects of cell planning for video broadcasting over WiMAX communication system [3], [1]. There are attractive growth rates in WiMAX subscriber base and equipment revenues, in market studies last years. 133 million subscribers will be supported at the end of 2012 [2].

WiMAX technology is one of the access technologies that enable transmission of IPTV services.. The QoS for delivering IPTV services depends especially on network performance and bandwidth [4]. The generic network topology of the IPTV application over WiMAX is shown in Fig. 1.

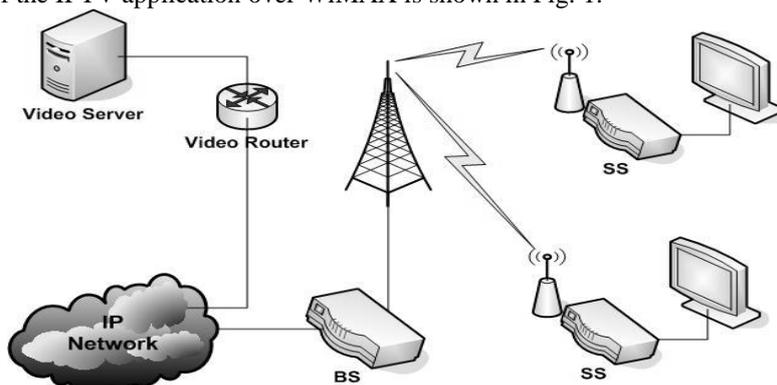

**Fig. 1**: System Model for IPTV Application [4]

Scalable Video Coding (SVC) has achieved significant improvements in coding efficiency with an increased degree of supported scalability relative to the scalable profiles of prior video coding standards. Transmitting SVC encoded videos over WiMAX networks is an effective solution which solves many of the video transmission problems over these networks [5].

IEEE 802.16 Medium Access Control (MAC) specifies five types of QoS classes are: Unsolicited grant services (UGS), which are designed to support constant bit rate services, such as T1/E1 and VoIP; Real-time polling services (rtPS), which are used to support real-time variable bit rate services such as MPEG video and VoIP; Non-real-time polling services (nrtPS), which are used to support non-real-time variable bit rate services such as FTP; Best-effort services, which enable to forward packets on a first-in-first basis using the capacity not used by other services; Extended real-time variable rate (ERT-VR) service, which is defined only in IEEE 802.16e, and is designed to support real time applications, that have variable data rates but require guaranteed data rate and delay [6].

Path loss is an unwanted introduction of energy tending to interfere with the proper reception and reproduction of the signals during transfer from transmitter to receiver. This environment between the transmitter and receiver in a wireless communication system has an important effect on the performance and to maintain QoS of the system. The path loss is an important element which must be kept within a predefined range in order to get the expected performance of system. The aim of this study is to investigate the performance of QoS of video streaming over fixed WiMax network under different terrain environments. The main objectives of this research are: To study the importance of IPTV (VoD), and to identify the factors affecting the VoD performance. Further, the study aims to simulate the WiMAX network with different propagation models in order to investigate and analyze the behavior and performance of these models.

The rest of this paper is organized as follows: Section II describes our system. Video traffic characteristics are described in Section III. Section IV describes simulation parameters. Simulation results and analysis was obtained in Section V. The study concludes in Section VI.

## II. SYSTEM DESCRIPTION

Our system was used is a fixed WiMAX system utilize OFDM over a 2.5 GHZ frequency, the system utilizes TDD with 5 MHz channel bandwidth that provisions 512 subcarriers allocated. The base station was configured to transmit power at 35.8 dBm which about 3.8 watts with 15 dBi gain antenna. Each client station transmit power was configured to use 33 dB which around 2 watts of transmit power over the 5MHz channel bandwidth using 14 dBi gain antennas. Moreover, the common propagation models for path loss used in this system namely: Free Space path loss model, and the Outdoor to Indoor and Pedestrian Environment model. Free Space path loss model was employed with a conservative terrain model with moderate to heavy tree density, representing rural environments, and has highest path loss with ignoring the shadow fading. Outdoor to Indoor and Pedestrian Environment model was employed with a conservative terrain model with moderate to heavy tree density, representing rural environments and has highest path loss, with 10dB shadow fading.

## III. VIDEO TRAFFIC CHARACTERISTICS

This section discusses some issues that come up when have been needed to deploy video in any network. These issues are characterizing the nature of video traffic, and QoS requirements. The QoS requirement is very important for deploying IPTV and VoD as realtime services. It is affected by packet loss, packet reordering, packet faults latency, packet duplication,and delay jitter. Some of the key metrics that define the level of service sustainable by a network upon a VoD deployment are as follows:

**Packet End-to-end delay:** Small amount of delay does not directly affect the Quality of Experience (QoE) of IPTV. While the delay large than 1 second may result a much worse QoS toward end-user experience. The delay for one way must be less than 200ms. On the other hand, the end-to-end delay more than 400ms was considered to be unacceptable [7].

**Packet loss:** Average number of packets lost compared to send a packet per second. Packet lost in the video processing layer is used for comparison purpose. There are some standard packet loss ratios given by ITU-T for classifying IPTV services. The excellent service quality of packet loss ratio is than $10^{-5}$ and the poor service quality of packet loss ratio is between $2*10^{-4}$ and 0.01 [7], a packet loss ratio above 1% is unacceptable.

**Jitter:** Calculated as the signed maximum difference in one way delay of the packets over a particular time interval [8]. Generally, jitter is defined as the absolute value of delay difference between selected packets. The jitter delay for one way must be less than 60ms on average and less than 10 ms in ideal [2].

**Throughput:** the rate at which a computer or network sends or receives data. It therefore is a good measure of the channel capacity of a communication link and connections to the internet are usually rated in terms of how many bits they pass per second (bit/s). The minimum end-to-end transmission rate acceptable for video is between 10 kbps and 5 Mbps [2].

## IV. SIMULATION

This section describes the simulation model adopted for analyzing the effect of Video on Demand (VoD) over the WiMAX networks .The simulation was performed to evaluate the performance of VoD over the WiMAX networks. Our simulation approach used the popular MIL3 OPNET Modeler simulation package1, Release14.0.A [9]. We used the OPNET Modeler to facilitate the utilization of in-built models of commercially available network elements, with reasonably accurate emulation of various real life network topologies.

The Wireless Deployment Wizard of OPNET is used to deploy a WiMAX network, with multiple subscriber stations in the range of a base station as shown in Fig. 2. The base stations are connected to the core network by an IP backbone. There is a server backbone containing the voice server which is configured as the video server. The IP backbone, server backbone together represent the service provider company network. The cell radius is set to 1 kilometres. The remaining key network configuration parameters in OPNET are summarized as in Table 1.

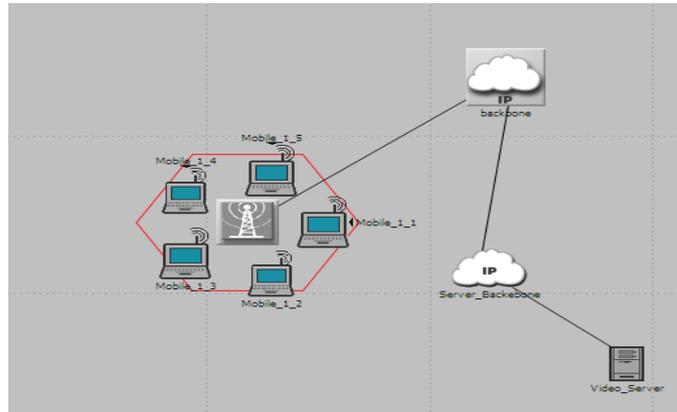

**Fig. 2**: OPNET Model of WiMAX Network

**Table 1**: Network Configuration Details

| Network | Fixed WiMAX Network |
|---|---|
| **Cell Radius** | 1 Km |
| **No. of Base Stations** | 1 |
| **No. of Subscriber Stations** | 5 |
| **IP Backbone Model** | Ip32_cloud |
| **Video Server Model** | PPP_sever |
| **Link Model (BS-Backbone)** | PPP_DS3 |
| **Link Model (Backbone-server Backbone)** | PPP_SONET_OC12 |
| **Physical Layer Model** | OFDMA 5 MHz |
| **Traffic Type of Services** | Streaming Video |
| **Application** | Real Video streaming |
| **Scheduling** | rtPS |

Video streaming over wireless networks is challenging due to the high bandwidth required and the delay sensitive nature of video than most other types of application. As a result, a video traces of Tokyo Olympics codec by SVC was used in our simulation. This traffic was obtained from Arizona State University [10], [11], with a 532x288 from resolution, Group of Picture (GOP) size is selected as 16 for this video for all codes, and encoded with 30 frames per seconds (fps).

Different scenarios have been designed in our simulation with different propagation models for the SVC video streaming over WiMAX to investigate the combined effect of these propagation models on the Quilty of Service (QoS) for video streaming over WiMAX in terms of: throughput, packet loss, path loss, and packet end-to-end delay.

## V. SIMULATION RESULTS AND ANALYSIS

Snapshots of OPNET simulation results for deploying VoD services over WiMAX with different propagation models was presented in this section. The goal of our simulation was to test the deployment, and to analysis the performance metrics IPTV (VoD) over WiMAX networks in different propagation models. The simulation results of our model are averaged across the 74 minute movie. Average values of end-to-end delay, path loss, packet loss, and throughput are used for the analysis in all the Figures.

In Fig. 3, we illustrate the corresponding streaming video end-to-end delay. As can be noted, that the video packet end to end delay of the network with free space as the path loss model is the lowest. However, the packet end-to-end is the highest for outdoor to indoor and pedestrian. This is due to the fact that as the density of the building structures, the number of times the signal gets obstructed and reflected, trees or mountains increases, and diffraction.

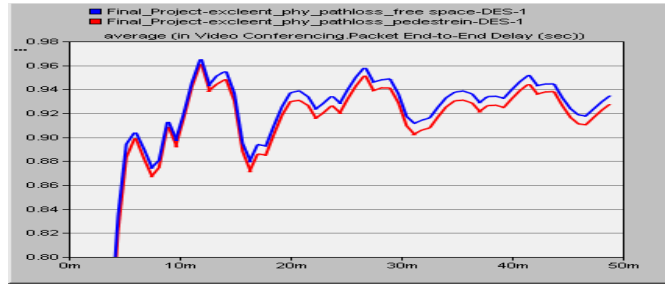

**Fig. 3**: Averaged End-to-End packet delay

The results of throughput for video streaming that is used in this simulation as shown in Fig. 4. It is observed that the throughput of the network with free space path loss model is the highest and the same for outdoor to indoor and pedestrian is the lowest for video traffic. This is due to the fact that as the density of obstacles increases, the Line of Sight (LOS) gets affected. As the Line of Sight (LOS) between the transmitting and receiving nodes decreases, it causes delays. Thus the number of times the signal gets obstructed and reflected is increasing. This results in increasing attenuation and diffraction due to the building structures, trees or mountains, also this result caused in packet loss.

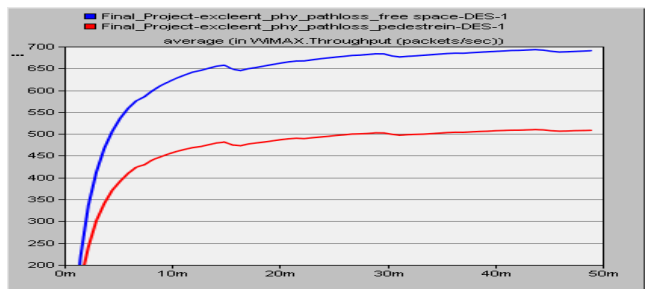

**Fig. 4**: Average throughput

Fig. 5 shows the path loss for Fress Space model that is around 100dB. Whereas, the path loss for the Outdoor to Indoor and Pedestrian is around 145dB . This indicates that the path loss for outdoor to indoor and pedestrian is the highest and the same for free space is the lowest. This is due to the fact that the path loss value varies with the amount of reflection in the communicating path.

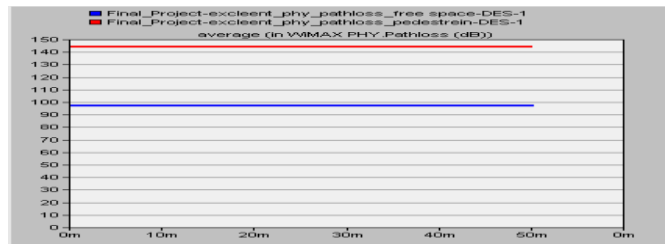

**Fig.5**: Path loss

The dropped packet rates by the PHY layer for the WiMAX Subscriber Station are shown in Fig. 6, which exhibits a much higher loss rate in Outdoor to Indoor and pedestrians. The low SNR for the subscriber station is a major contributor to the high packet loss rate.

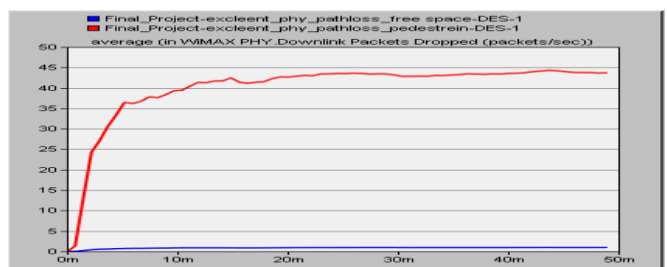

**Fig. 6**: packet lost

## VI. CONCLUSION

This study explores the performance analysis of IPTV over WiMAX broadband access technology. Its aim is to evaluate the effect of various path loss models, on the basis of average throughput and packet end to end delay and path loss. The OPNET Modeler is used to design and characterize the performance parameters of Tokyo Olympics video streaming under different terrain environment. The simulation results indicate that, free space path loss model is a basic path loss model with all other parameters related to terrain and building density set as constant. Furthermore, the streaming video content has been modeled as unicast traffic while multicast video traffic may have yielded better performance. This work has limitations to certain assumptions like: Station transmit power, distance between base station and subscriber station, subscriber station was configured as fixed not support mobility, station antenna gain, carrier operating frequency and channel bandwidth.